\documentstyle[aps,prl,epsfig,multicol]{revtex}
\makeatletter
\begin{document}
\title{Electron-hole generations: A numerical approach to
interacting fermion systems}
\author{Richard Berkovits}
\address{Minerva Center and Department of Physics,
Bar-Ilan University, Ramat-Gan 52900, Israel}
\date{January 21, 2002, version 2.0}
\draft
\maketitle
\begin{multicols}{2}[%
\begin{abstract}
A new approach, motivated by Fock space localization,
for constructing a reduced many-particle Hilbert space is
proposed and tested. The self-consistent Hartree-Fock (SCHF) approach
is used to obtain a single-electron basis from which the many-particle
Hilbert space is constructed. For a given size of the truncated many
particle Hilbert space only states with the lowest number of
particle-hole excitations are retained and exactly diagonalized. 
This method is shown to be
more accurate than previous truncation methods, while there is no
additional computational complexity.
\end{abstract}
\pacs{PACS numbers: 72.15.Rn,71.55.Jv,71.10.Fd}
]

There has been a growing interest in the physics of
low temperature disordered many-particle
systems such as the two dimensional ``metal-insulator'' transition
\cite{abrahams01}, electron dephasing due to interactions \cite{aleiner99}
and transport through quantum dots\cite{alhassid00}. Unfortunately these
problems are very difficult to treat analytically since neither disorder nor
interactions may be considered as a small perturbation. An exact
numerical treatment is limited to small systems due to the large size
of the many-particle Hilbert space. For a system
of $m$ single particle orbitals and $n$ electrons, the
many-particle Hilbert space size $M=\left(^{m}_{n}\right)$.

One popular approximation method for low-lying eigenstates (especially the
ground state) is to reduce the size of the many-particle Hilbert space,
i.e., to limit the number of many-particle basis
states spanning the Hilbert space. One strategy is to include
in the many-particle basis states only single-electron states which are lower
in their non-interacting energy than a certain threshold. This will
limit the number of single-electron states $m_R<m$, reducing the
number of many-particle states to $M_R=\left(^{m_R}_{n}\right)$.
This procedure works well in the strongly localized 
regime\cite{efros95,talamantes96},
but fails for weaker disorder. As demonstrated bellow,
a similar strategy
using Hartree-Fock single-electron states, works better in
the diffusive regime.

One might also reduce $M$ by imposing 
a global constraint on the many-particle states.
For example, the many-particle basis states may be composed of $M_R<M$ 
Fock determinants of single-electron states for which the total
non-interacting energy or
the expectation value of the full interacting Hamiltonian
is smaller than
some threshold \cite{song00}.
An even better
procedure is to use 
Hartree-Fock single-electron states to
build the many-particle basis, while picking the $M_R$ states with
the lowest values of the
expectation value of the interacting Hamiltonian\cite{vojta98,vojta01}. 
All these
methods suffer from
the drawback of forcing one to go over all $M$ basis states in order to chose
$M_R$ of them according to the global criteria. Since $M$ grows
exponentially, this is not
a practical solution. Nevertheless, it is possible to overcome this snarl by
sampling only part of all the $M$ states, for example by a
Monte Carlo procedure\cite{vojta98}.

In this paper we propose a different way to truncate the many-particle
basis which gives a more accurate evaluation of the low-lying
eigenvalues and eigenvectors for the same truncated basis size $M_R$
than the previous methods. Moreover, the truncation criteria is not global
and does not require evaluating some criteria for all the many-particle basis
states, i.e., the truncation has the same computational complexity as the
single-electron criteria truncation but nevertheless produces more accurate 
results than the complicated global criteria methods. 
Our truncation method is based on the fact that the interaction
operator is a two-body operator. Thus, only many-particle states which
differ by at most two e-h (electron-hole) 
pairs can be coupled by the interaction
operator. This fact is well known in the context of nuclear physics
\cite{french70,bohigas71},
and has recently been brought to the forefront
of disordered interacting systems by the concept of Fock space localization
by Altshuler,  Gefen,  Kamanev and Levitov
\cite{altshuler97}. We therefore propose to construct the many-particle
basis in the following way: First a many-particle
basis state is constructed as a determinant from the $n$ lowest-lying
Hartree-Fock single-electron states. Then all the many-particle
basis states differing from the first state by up to $k$ e-h
pairs are included, resulting in a truncated basis of size
\begin{eqnarray}
\label{eq:mr}
M_R=
\sum_{j=0}^k \left(^{m-n}_j\right) \left(^{n}_j\right).
\end{eqnarray}

From the point of view of the Fock space localization postulate
\cite{altshuler97}
this truncation method has the potential for being accurate for finite systems
provided $k$ is
larger the Fock space localization length $\xi_F$, since the
average contribution of a state in the $k$-th generation
to the exact eigenstate of the system is proportional
to $\exp(-k/\xi_F)$. Thus, the contribution of the state of the $k$-th
e-h generation to the exact eigenstate is proportional to
$\left(^{m-n}_k\right) \left(^{n}_k\right) \exp(-k/\xi_F)$, i.e.,
as long as $1/\xi_F > \ln((m-n)n)$ the weight of each consecutive generation
falls of exponentially. Although
there is an ongoing debate regarding the critical energy at which
the transition occurs
and whether the transition is  abrupt or more gradual
\cite{mirlin97,silvestorov97,georgeot97,flambaum97,berkovits98,mejia98,leyronas99,leyronas00},
it is nevertheless clear that
for low-lying excitations the contribution of large $k$ states will
fall off strongly.

We illustrate the method for an interacting disordered many-particle
two dimensional tight-binding Hamiltonian:
\begin{eqnarray}
H &=& H_{0} + H_{\rm int} \nonumber \\
H_{0} &=&
\sum_{k,j} \epsilon_{k,j} n_{k,j}
- V \sum_{k,j} [a_{k,j+1}^{\dag} a_{k,j} +
a_{k+1,j}^{\dag} a_{k,j} + h.c. ] \nonumber \\
H_{\rm int} &=& U \sum_{k,j} n_{k,j} n_{k \pm 1,j \pm 1},
\label{hamil}
\end{eqnarray}
where $\{k,j\}$ denotes a lattice site, $a_{k,j}^{\dag}$
is an electron creation operator, the number operator is
$n_{k,j}=a_{k,j}^{\dag} a_{k,j}$,
$\epsilon_{k,j}$ is the site energy chosen randomly between $-W/2$ and
$W/2$ with uniform probability and $V$ is a constant hopping matrix
element.
Only nearest neighbor interactions $U$ are taken into account.
This is reasonable in the metallic regime where screening is
good.

Systems composed of a $n=4,6,10$ electrons residing on $m=16,15,24$ sites of
$4 \times 4,5 \times 3,6 \times 4$ 
lattice (i.e., $M=1820,5005,1961256$),
were considered. The latter size is at the limit of
the possibility for exact diagonalization using the Lanczos method
for current computer technology, while the former sizes are typical
for standard exact diagonalization procedures.
Disorder strength of $W=8V$ for the $4 \times 4$ and $5 \times 3$ lattices and
$W=5V$ for the $6 \times 4$ lattice
were chosen so that the
single particle localization length is larger than the system
size, while the
single-electron states follow the Gaussian orthogonal ensemble (GOE).

We carry out exact diagonalization of the many-particle
Hamiltonian using standard procedures for the smaller systems
and the Lanczos method for the larger ones, obtaining the
exact many-particle eigenvalues
$E_{\alpha}$ and eigenvectors $|\alpha\rangle$,
where $\alpha=0,1,2$ for the ground state, the first and second
excitation. 

In order to obtain the single-electron states we diagonalize the single
electron part $H_0$ in the Hamiltonian given in Eq. (\ref{hamil}),
obtaining the single-electron
eigenvectors $|\varphi_i\rangle$ and eigenvalues $\varepsilon_i$.
For the Hartree-Fock single-electron wavefunctions
we solve the
self consistent Hartree-Fock (SCHF) Hamiltonian:
\begin{eqnarray}
\label{eq:schf}
H_{HF}= H_{0} +
\sum_{k,j} n_{k,j} U \langle n_{k \pm 1,j \pm 1} \rangle_0 \\ \nonumber
-\sum_{k,j} a_{k,j}^{\dag} a_{k \pm 1,j \pm 1} U
\langle a_{k \pm 1,j \pm 1}^{\dag} a_{k,j} \rangle_0,
\end{eqnarray}
where $\langle\ldots\rangle_0$ denotes an average on the ground
state which is calculated self-consistently. The SCHF single-electron
eigenvectors $|\psi_i\rangle$ and eigenvalues $\zeta_i$ are
obtained through a self consistent diagonalization of the
Hartree-Fock Hamiltonian.

Any many-particle wavefunction $|\alpha\rangle$ may be expressed in terms of
a combination of Fock determinants composed of the single-electron
(or Hartree-Fock single-electron) eigenvectors in the following way:
\begin{eqnarray}
\label{fock}
|\alpha\rangle = \sum_{i_1,i_2,\ldots,i_N}
C_{i_1,i_2,\ldots,i_n} c_{i_n}^{\dag}\ldots c_{i_2}^{\dag} c_{i_1}^{\dag}
|{\rm vac}\rangle,
\end{eqnarray}
where the sum is over all the possible combinations of $N$ states out of
the $M$ possible ones for which
$ m \geq i_n>i_{n-1}>\ldots>i_2>i_1 \geq 1$, $c_i^{\dag}$ is the creation
operator of the $i$-th single-electron (or Hartree-Fock single-electron) state,
$|{\rm vac}\rangle$ is the vacuum state.

Let us now define more precisely the various truncation methods which
we have discussed above. The first method (which we shall name
single-electron energy truncation)
is to include
in the many-particle basis states only single-electron states for which
$\varepsilon_i<\varepsilon_R$. This method will
reduce the number of single-electron states to $m_R<m$.
Usually $\varepsilon_R$ is chosen according to the maximum
$m_R$ desired.
The $M_R=\left(^{m_R}_{n}\right)$ many-particle basis
will include the states:
\begin{eqnarray}
\label{basis}
|\beta\rangle = c_{i_n}^{\dag}\ldots c_{i_2}^{\dag} c_{i_1}^{\dag}
|{\rm vac}\rangle,
\end{eqnarray}
for which $m_R \geq i_n>i_{n-1}>\ldots>i_2>i_1 \geq 1$. The variant of this
method 
using the single-electron Hartree-Fock states for which
$\zeta_i<\zeta_R$ will be called the
single-electron Hartree-Fock energy truncation. 
The basis in this case composed of the states described
in Eq. (\ref{basis}) in which now $c_i^{\dag}$ is the creation
operator or the $i$-th Hartree-Fock single-electron state.

The truncated Hamiltonian then may be diagonalized using
standard diagonalization procedures as long as $M_R$ is
not too big ($M_R\sim 10000$). For larger sizes one can
use the Lanczos method as long as the matrix remains sparse
enough. The sparseness of the many-particle matrix is
roughly $S\sim n^2(m_R-n)^2/\left(^{m_R}_{n}\right)$. Thus,
as long as $m_R$ is not to close to $n$ the Lanczos method
may be useful in obtaining the low-lying excitations.

For the methods based on truncation according to a global criteria,
one (in principal) should construct all possible basis states
$|\beta\rangle$ for which $m \geq i_n>i_{n-1}>\ldots>i_2>i_1 \geq 1$.
Then only states which correspond to some global criteria,
for example $\langle\beta|H|\beta\rangle < E_R$, are to be included
in the basis. Again this method may use either $|\varphi_i\rangle$
or $|\psi_i\rangle$ as the single-electron states from which the
many-particle
basis states $|\beta\rangle$ are composed. This method will be named
global energy truncation. In this case the sparseness
of the matrix depends on the global criteria, and for a small $M_R$ the
matrix is not necessarily sparse \cite{vojta98}.

In the e-h generation truncation method the basis states
are composed as follows: A basis state $|\beta_0\rangle =
c_{n}^{\dag}\ldots c_{2}^{\dag} c_{1}^{\dag} |{\rm vac}\rangle$
is chosen. Then all states which may be created from this
state by up to $k$ applications of an e-h pair
creation operator are added to the basis, i.e.,
\begin{eqnarray}
\label{gen}
|\beta\rangle = \Pi_{j=1}^k
\left(c_{l_j}^{\dag} c_{i_j} \right)
|\beta_0\rangle,
\end{eqnarray}
where $m \geq l_1 \ne l_2 \ne \ldots l_k>n$ and
$n \geq i_1 \ne i_2 \ne \ldots i_k \geq 1$. The total
number of basis states
for up to $k$ e-h generations is given in Eq. (\ref{eq:mr}).
The sparseness of a specific e-h generation $j$ is given by
$S_j \sim (n-j)^2(m-n-j)^2/\left(^{m-n}_{j-2}\right) \left(^{n}_{j-2}\right)$,
and the many-particle matrix sparseness will be determined mainly
by the higher generations. Thus, the higher the number of
e-h generations included, the sparser the matrix becomes.

\begin{figure}\centering
\epsfxsize12cm\epsfbox{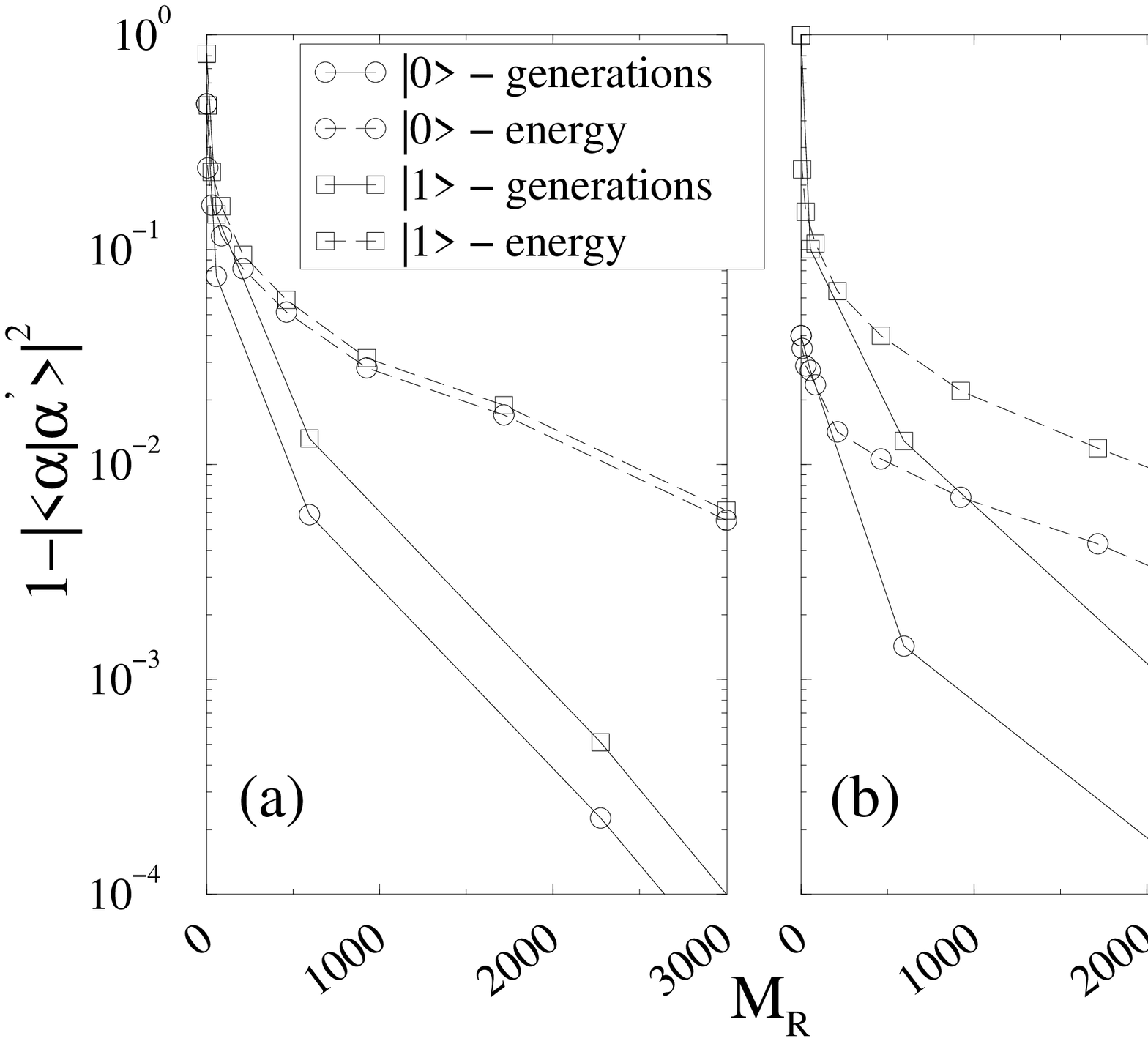}
\caption{The discrepancy between the exact eigenstate $|\alpha\rangle$
and the one $|\alpha \prime\rangle$
obtained via a truncated many-particle basis
for $n=6$ electron on $m=15$ sites of a $5 \times 3$ lattice (i.e., $M=5005$)
for $U=2V$.
The discrepancy is presented
as a function of the size of the truncated basis $M_R$, the state (
the ground state $|0\rangle$ (circles);
first excited state $|1\rangle$ (squares))
and the method of truncation. (a) The single-electron energy truncation
(dashed line), generation truncation method (full line).
(b) As in (a) but  the single-electron states are replaced by
Hartree-Fock states.
}
\label{fig1}
\end{figure}

In the following we shall compare the accuracy of the different
methods. First we compare the accuracy of the
single-electron energy truncation
to the e-h generation truncation method by applying both of them
to the $5 \times 3$ system.
In Fig. \ref{fig1} a comparison of the overlap between
the approximated wave function $|\alpha' \rangle$
obtained using the truncated basis $\{|\beta\rangle\}$
and the exact wave function $|\alpha\rangle$ is presented.
The results were averaged over $10$ different realizations of
disorder. For the single-electron energy truncation we have chosen
$\varepsilon_R$ for each realization such that values of
$m_R=6,7,8,9,10,11,12,13,14$ and
$M_R=1,7,28,84,210,462,924,1716,3003$ correspondingly are obtained.
This is compare with the results of the generation truncation
of $k=0,1,2,3,4,5$ and $M_R=1,55,595,2275,4165$. Both cases of
single-electron energy truncation (Fig. \ref{fig1}a) and
single-electron Hartree-Fock energy truncation (Fig. \ref{fig1}b)
were considered. It can be seen that for roughly the same size of the
truncated many-particle basis $M_R$ the generation truncation is
better by up to an order of magnitude
for either the single-electron or the Hartree-Fock
single-electron states. It is also interesting to note that
using Hartree-Fock states does not always improve the accuracy.
While for the ground state the Hartree-Fock single-electron states
give a better result in any truncation method than the non-interacting
single-electron states, for the first excited state this is so only
as long as $M_R$ is small compared to $M$.

\begin{figure}\centering
\epsfxsize12cm\epsfbox{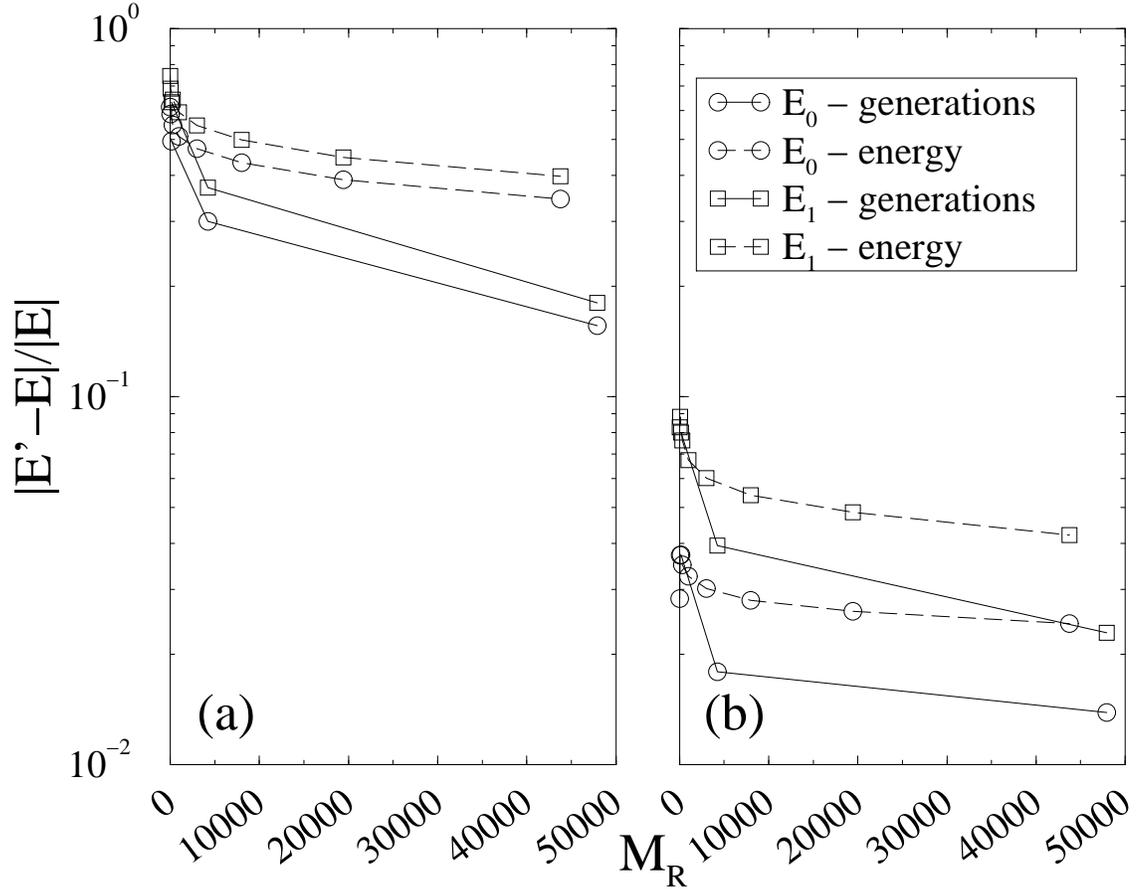}
\caption{The discrepancy between the exact eigenvalue $E_{\alpha}$
and the one ${E_{\alpha}} \prime$ obtained via a truncated many-particle basis
for $n=10$ electron on $m=24$ sites of a $6 \times 4$ lattice
(i.e., $M=1961256$) for $U=3V$. The discrepancy is presented
as a function of the size of the truncated basis $M_R$, the energies (
the ground state $E_0$ (circles);
first excited state $E_1$ (squares))
and the method of truncation. (a)  The single-electron energy truncation
(dashed line), generation truncation method (full line).
(b) As in (a) but the single-electron states are replaced by
Hartree-Fock states.
}
\label{fig2}
\end{figure}

In Fig. \ref{fig2} we check the accuracy of the different truncation
methods for the eigenvalues of a much larger $6 \times 4$ system.
The results were averaged over $100$ realizations.
For the single-electron energy truncation we have chosen
$\varepsilon_R$ for each realization such that values of
$m_R=10,11,12,13,14,15,16,17,18$ and
$M_R=1,11,66,286,1001,3003,8008,19448,43758$ correspondingly are obtained.
This is compared with the results of the e-h generation truncation
of $k=0,1,2,3$ and $M_R=1,141,4236,47916$. The most obvious observation
is that using Hartree-Fock states in any truncation scheme is much
better than using non-interacting single-electron states. This stems from the
fact that for this large system $M_R \ll M$. Again, for approximately
the same many-particle basis size $M_R$ the generation truncation method
is more accurate.

\begin{figure}\centering
\epsfxsize12cm\epsfbox{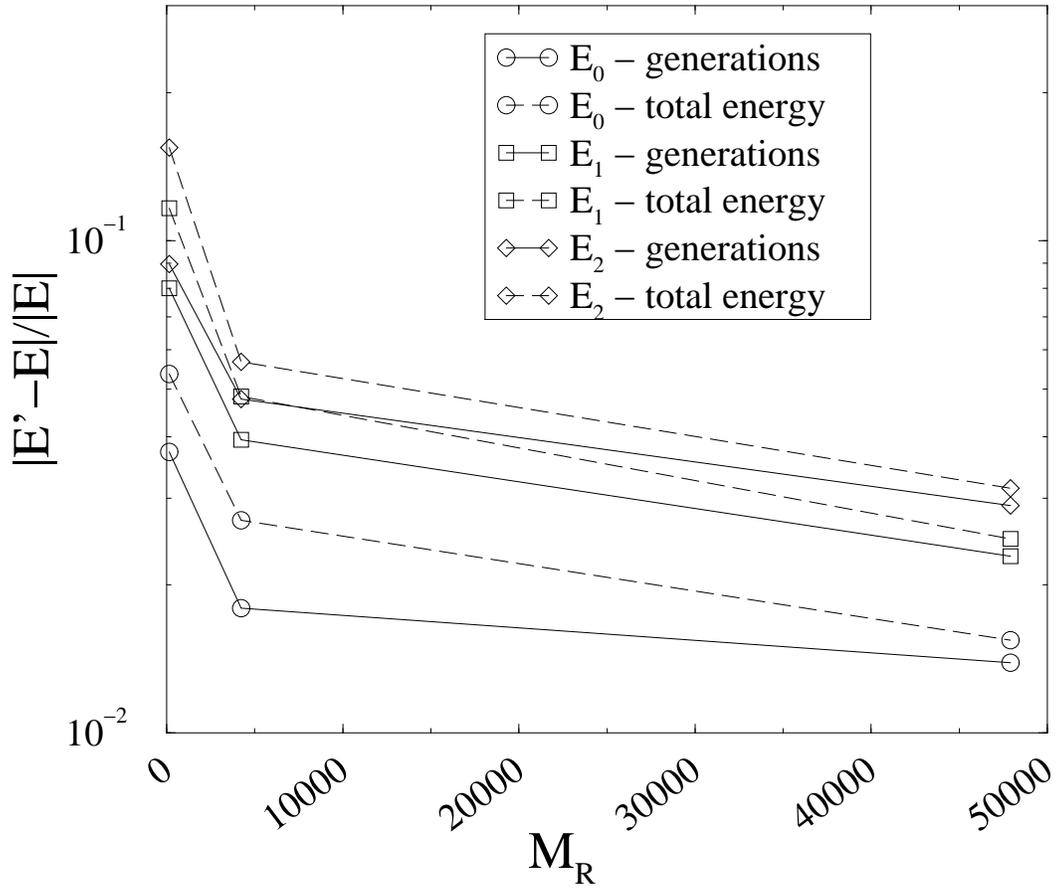}
\caption{The discrepancy between the exact eigenvalue $E_{\alpha}$
and the one ${E_{\alpha}} \prime$ obtained via a truncated many-particle basis
for $n=10$ electron on $m=24$ sites of a $6 \times 4$ lattice
(i.e., $M=1961256$) for $U=3V$. The discrepancy is presented
as a function of the size of the truncated basis $M_R$, the energies (
the ground state $E_0$ (circles);
first excited state $E_1$ (squares); second
excited state $E_2$ (diamonds))
and the method of truncation.
The dashed line corresponds to the global energy truncation, while
the full line
corresponds to the generation truncation.
In both cases Hartree-Fock single-electron states were used.
}
\label{fig3}
\end{figure}

A more stringent test for the generation truncation method is its comparison
to the global energy truncation. In Fig. \ref{fig3} the accuracy of the
three lowest eigenvalues for the generation truncation and
global energy truncation method are compared for a $6 \times 4$ system.
The energy generation truncation
after $k=1,2,3$ generations (i.e., $M_R=141,4236,47916$) is compared
to a global energy truncation of the same basis size when the single level
states in both cases are composed of Hartree-Fock states.
For both methods the accuracy is better for the ground state than for higher
excitations. In all cases the generation truncation performs better at a given
$M_R$ than the global energy truncation. Moreover, the advantage of
the generation truncation is larger for the lower values of $M_R$.

\begin{figure}\centering
\epsfxsize12cm\epsfbox{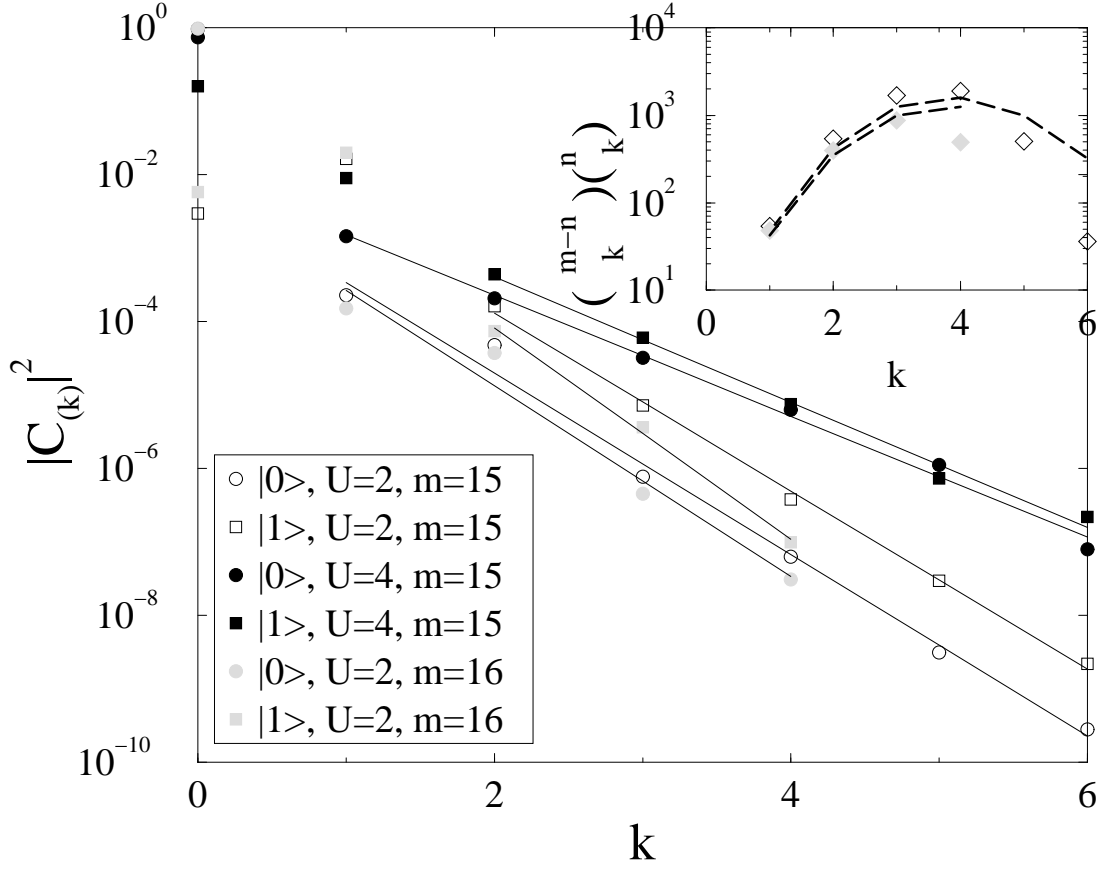}
\caption{The average weight of the exact ground state $|0\rangle$ (circles) and
first excited state  $|1\rangle$ (squares) per a k-th generation 
$\langle C_{(k)}^2 \rangle$ for
$n=6$ electron on a $5 \times 3$ ($m=15$)
lattice at $U=2V$ (white symbols) and
$U=4V$ (filled symbols), and  for
$n=4$ electron on a $4 \times 4$ ($m=16$) lattice at $U=2V$ (grey symbols).
All cases were averaged over $10$ realization of disorder.
A best fit to an exponential $\exp(-k/\xi_F)$
is indicated by the straight lines.
Inset: the number of
states per a k-th generation 
$\left(^{m-n}_k\right) \left(^{n}_k\right)$
is indicated by the diamonds (white symbols $n=6,m=15$, grey symbols 
$n=4,m=16$). 
The dashed line corresponds to the approximation
$((m-n)m)^k exp(-(m/2(m-n)n)k^2) / (k!)^2$.
}
\label{fig4}
\end{figure}

Let us now examine whether one can formulate a quantitative estimation
for the contribution of different e-h generations.
The weight of a given Fock determinant
in the exact eigenvector is 
$|C_{i_1,i_2,\ldots,i_n}|^2$ (see Eq. (\ref{fock})). 
By averaging over all Fock determinants belonging
to a certain e-h generation $k$ one obtains the average weight
of a state in the  k-th generation $\langle C_{(k)}^2 \rangle$ 
depicted in Fig. \ref{fig4}. The 
average weight roughly falls off exponentially as a function of $k$,
$\langle C_{(k)}^2 \rangle \propto \exp(-k/\xi_F)$, where $\xi_F$ 
is the Fock space localization length. We find that
$\xi_F$ depends on the interaction strength 
$\xi_F(U=2V) \sim 0.4$ and $\xi_F(U=4V) \sim 0.7$
and has also a weaker dependence on $m$ and $n$.
This exponential behavior is in line with what is expected by the
Fock space localization scenario\cite{altshuler97}.
The number of states in the $k$-th e-h generation is equal to
$\left(^{m-n}_k\right) \left(^{n}_k\right)$, which for $k \ll n,m$
is proportional to $((m-n)n)^k \exp(-(m/2(m-n)n)k^2) / (k!)^2$
(see inset Fig. \ref{fig4}). Thus,
as long as $1/\xi_F > \ln[(m-n)n/(k+1)^2]$
the weight of each consecutive generation beyond the $k$-th generation
becomes exponentially smaller.

In conclusion, using the ideas of Fock space localization
we have proposed and tested a new truncation method for reducing the size 
of the many-particle Hilbert space. The method is based on retaining states
with the lowest number of electron hole excitations above the SCHF ground
states.
This method is shown to be more accurate 
than other widely used truncation methods, while requiring no
additional computational complexity. Refinements of the e-h generation 
truncation methods (e.g., adding a single-electron energy 
truncation condition) merit further study.

I would like to thank Y. Gefen and B. L. Altshuler for useful discussions and
the Israel Science foundation for financial support.

\end{multicols}

\vfill\eject

\vfill\eject

\vfill\eject

\end{document}